\documentclass[proceedings]{JHEP}
\usepackage{epsfig}
\usepackage{pstricks}
\usepackage{rotate}
\conference{Heavy Flavours 8, Southampton, UK, 1999}
\title{$ep$ Physics with Heavy Flavours}
\author{Felix Sefkow \\ 
        Physik-Institut der Universit\"at Z\"urich,
        CH-8057 Z\"urich, Switzerland \\
        Email: {\email{Felix.Sefkow@desy.de}}
}
\abstract{Recent results from H1 and ZEUS 
on heavy flavour production at HERA are reviewed.} 

\newcommand{\ra}{\rightarrow}
\newcommand{\ccb}{c\bar{c}}

\newcommand{\ftc}{F_2^c}
\newcommand{\pom}{{I\!\!P}}
\newcommand{\zpom}{z_{_{I\!\!P}}}

\begin{document}

\section{Introduction}
Heavy Flavour physics at HERA focuses on aspects related to 
production dynamics rather than weak decays and mixing angles.
Heavy quarks
-- like jets -- reflect the properties of hard sub-processes
in $ep$ interactions. 
Their mass provides the natural scale 
which allows the application of perturbative
methods in QCD and also testing the theory 
in regions where no other hard scale 
(like high transverse energy) is present. 
Furthermore, when probing the structure of matter, heavy quarks 
single out the gluonic content. 

For this talk topics from open charm and beauty production  
have been selected. 
There was no time to cover Onium production (for recent reviews, 
see~\cite{naroska}).
There is no tau physics included, either, although $\tau$ candidates have 
been seen at HERA, and interesting limits on lepton-flavour violating
lepto-quarks have been obtained~\cite{lq}. 

At HERA, 820 GeV protons (defining the ``forward'' direction) collide
head-on with 27~GeV electrons, 
yielding a centre-of-mass system (CMS) energy of $\sqrt{s}=$ 300 GeV. 
The standard textbook notation will be used to describe the kinematics 
of deep inelastic scattering (DIS): 
$Q^2$ denotes the four-momentum transfer squared, 
$x$ and $y$ are scaling variables related to the momentum fraction of the 
struck parton and to the inelasticity of the collision, respectively.
$W$ denotes the photon-proton CMS energy. 
The kinematic regime where the exchanged photon becomes qua\-si-real 
($Q^2\ra 0$) is called photo-pro\-duc\-tion.
Also in this region 
$W$ can be large 
and attains values an order of magnitude 
higher than in fixed-target experiments.

\section{Charm Production}

In most of the studies presented here, open charm is detected
in the ``golden'' decay channel $D^{\ast +} \ra D^0\pi^+ $ followed 
by $D^0\ra K^-\pi ^+$.%
\footnote{The charge conjugate is always implicitly included.} 
ZEUS also uses the channel $D^{\ast+}\ra(D^0\ra K^-\pi^-\pi^+\pi^-)\pi^-$,
and this year for the first time they have shown 
results using   
$D_s+\ra(\phi\ra K^+K^-)\pi^+$~\cite{zeusdsubs}.

\subsection{Deep inelastic scattering}

In QCD, $c$ and $b$ production in $ep$ collisions proceed mainly via 
the boson gluon fusion diagram shown in figure~\ref{fig:bgf}.
\FIGURE{
    \unitlength5mm
    \begin{picture}(10,9)(3,0)
      \put(3,0){
          \epsfig{file=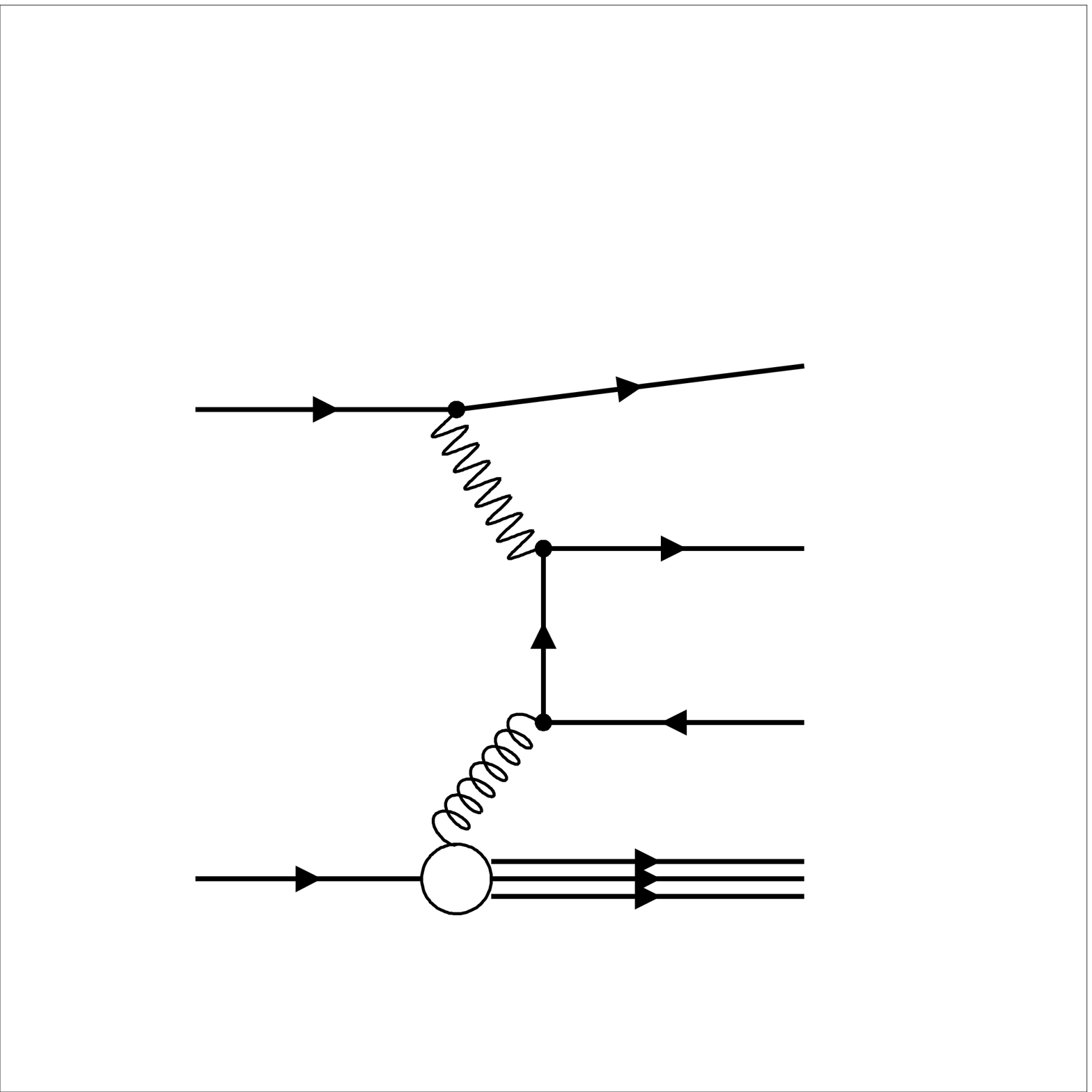,
bbllx=70.,bblly=70.,bburx=450.,bbury=400.,clip=,width=5cm}
}
\put(2.7,1){\Large $p$}
\put(2.7,7.3){\Large $e$}
\put(6,2.5){\Large\blue $g$}
\put(6,6){\Large $\gamma$}
\put(13,3){\Large\red $\bar{c},\bar{b}$}
\put(13,5.5){\Large\red $c,b$}
\end{picture}
\caption{\label{fig:bgf} Boson gluon fusion.}
}

Therefore, charm production has already in the past offered a way 
to determine the gluon density in the proton~\cite{oldcharmglue}.
The process has been calculated in Next-to-Leading order (NLO) QCD 
in the so-called Three Flavour $\overline{MS}$ scheme
for photo-production~\cite{frixione} and DIS~\cite{disnlo}.  
In the HVQDIS program, the charm hadronization into $D^{\ast}$ me\-sons
is modeled using a Peterson fragmentation function~\cite{peterson}
with the parameter $\epsilon_c = 0.035$ as determined in $e^+e^-$ 
annihilation~\cite{opal0.035}.
\FIGURE{
    \unitlength1cm
\epsfig{file=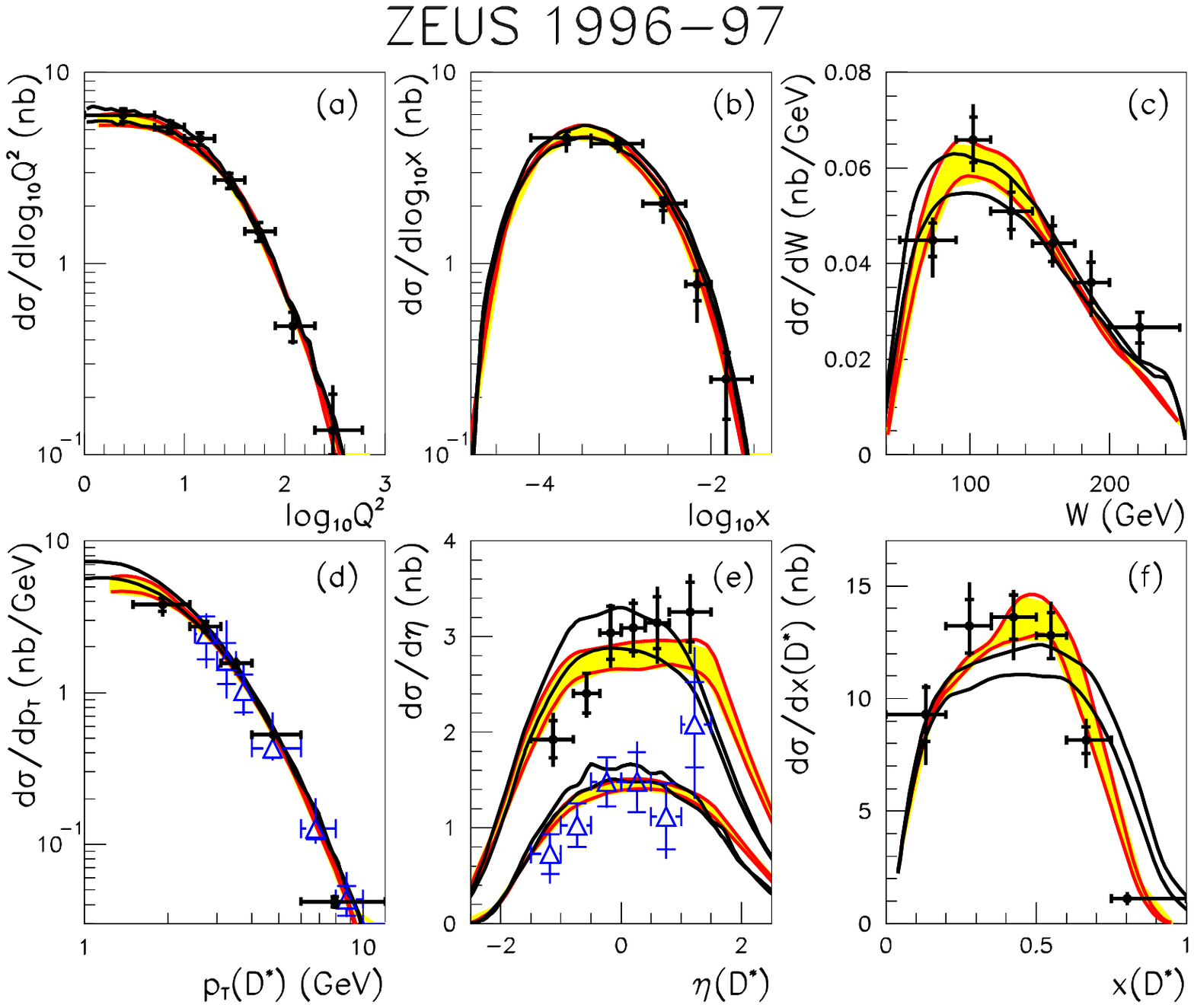,width=11cm}
\caption{\label{fig:disXs} $D^{\ast}$ cross sections in DIS,
compared with NLO QCD, using for fragmentation 
a Peterson type (open band) 
or a LO MC model (shaded band).
The widths of the bands correspond to a charm mass variation 
between 1.3 and 1.5 GeV.}
}

Differential $D^{\ast}$ cross sections in the experimentally accessible 
DIS range have been measured by H1~\cite{glue} and ZEUS~\cite{zeusf2c}.
The ZEUS data are shown in figure~\ref{fig:disXs} and compared to 
the HVQDIS calculation. 
(The open points in (d) and (e) are results from the $K4\pi$ channel.)
A Leading-Order Monte Carlo fragmentation approach 
based on the JETSET program also has been used by ZEUS.
As can be seen from the figure, the pseudo-ra\-pi\-di\-ty 
and the $x_D$ distribution%
\footnote{$x_D\equiv 2p^*(D^*)/W$ is the fractional momentum of the meson
in the $\gamma ^* p$ rest frame and strongly correlated with~$\eta$.}  
are quite sensitive to details of the fragmentation modeling, 
whereas the distributions of other kinematic variables, like $x$ and $Q^2$ are
barely affected.  
The migration of the mesons towards positive pseudo-rapidities 
in the Monte Carlo, due to color interactions 
between the charm quark and the proton remnant, 
has been called the ``beam drag'' effect~\cite{norrbin}. 

In general, there is good agreement with the NLO calculation. 
This gives the justification to the extrapolation to the full phase space, 
which is needed to extract $\ftc$, the charm contribution to the 
proton structure function.

In the Quark-Parton Model, the structure function $F_2$ is given
by the charge-weighted sum of the (anti-)quark densities,
$F_2=\sum_i e_i^2(q_i(x)+\bar{q}_i(x))$ and depends only on $x$.
The presence of gluons introduces ``scaling violations'', i.e.\ 
a dependence of the structure function on $Q^2$. The gluons
are also responsible for the observed steep rise of $F_2$ towards
low values of $x$, where the quark sea is being probed. 
\FIGURE{
    \unitlength1cm
\epsfig{file=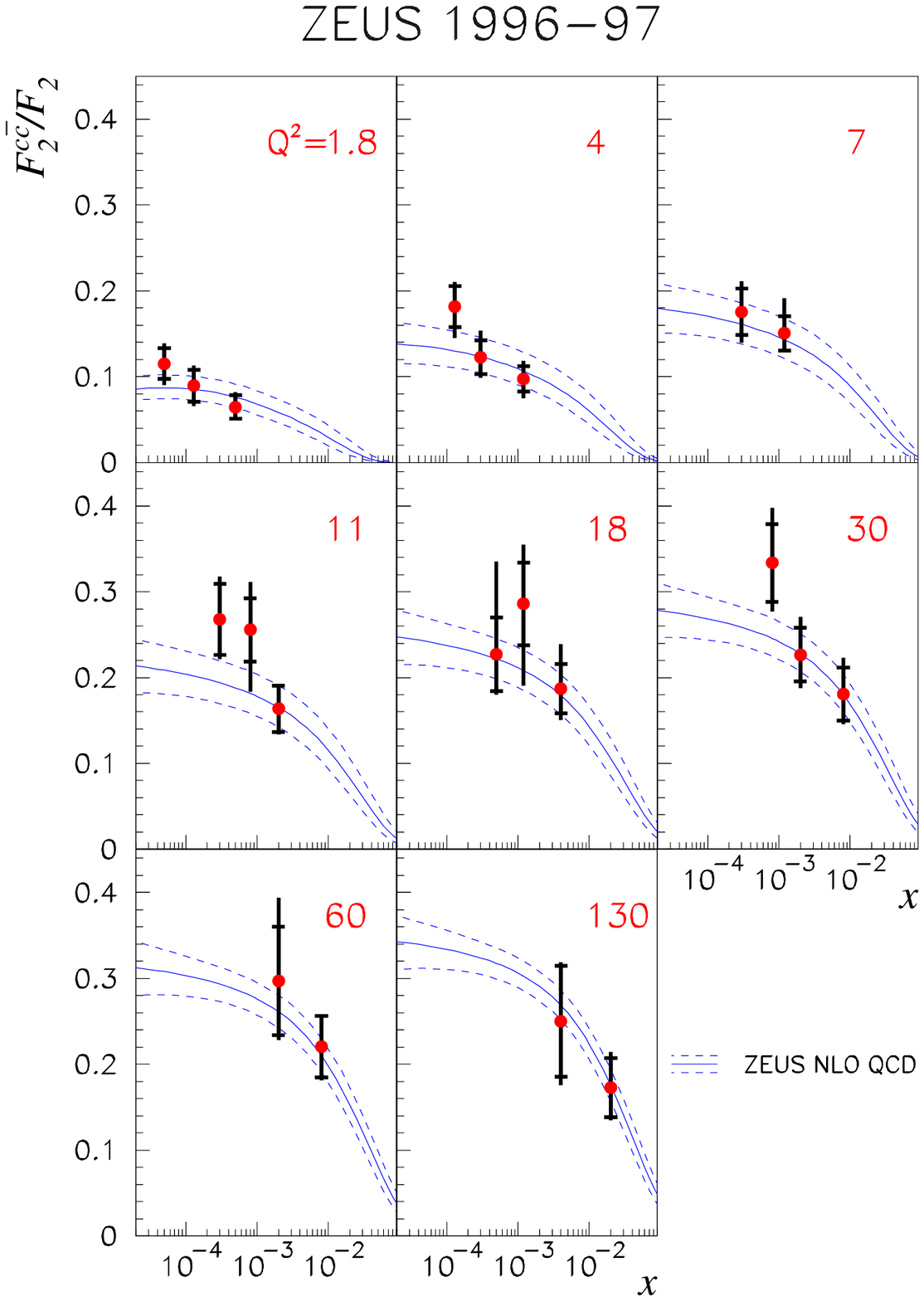,width=6.9cm}
\caption{\label{fig:f2c} Relative charm contribution to inclusive DIS.
(Extrapolation uncertainties of $\sim 20\%$ are not shown.)}
}

The relative charm contribution to inclusive DIS, quoted as $\ftc / F_2$,
is not constant, but rises with $Q^2$ and towards low $x$, as can be seen
in figure~\ref{fig:f2c}.
Since the process under study is gluon-induced, $\ftc$ reflects the gluon 
content of the proton even more pronouncedly than $F_2$. 
The most salient feature is that in the HERA regime the charm 
contribution is very large, 20 - 30\%, making the theoretical description 
of charm production an essential ingredient to the understanding of 
proton structure. 

H1 has performed a direct measurement of the gluon density in the proton
from charm production in DIS and in photo-production. 
The ki\-ne\-ma\-tics of the reconstructed $D^{\ast}$ meson and the scattered 
electron have been used to infer the incoming gluon momentum. 
The NLO programs have been used in order to correct for 
higher order processes and fragmentation by means of an un\-fol\-ding 
procedure. 
The results, shown in figure~\ref{fig:glue}, 
agree well with each other and with the gluon distribution 
determined indirectly through a QCD analysis of the scaling violations 
of inclusive structure function data. 
This demonstrates that the gluon density is universal, and 
indicates the success
of the NLO QCD description. 
\FIGURE{
    \unitlength1cm
\epsfig{file=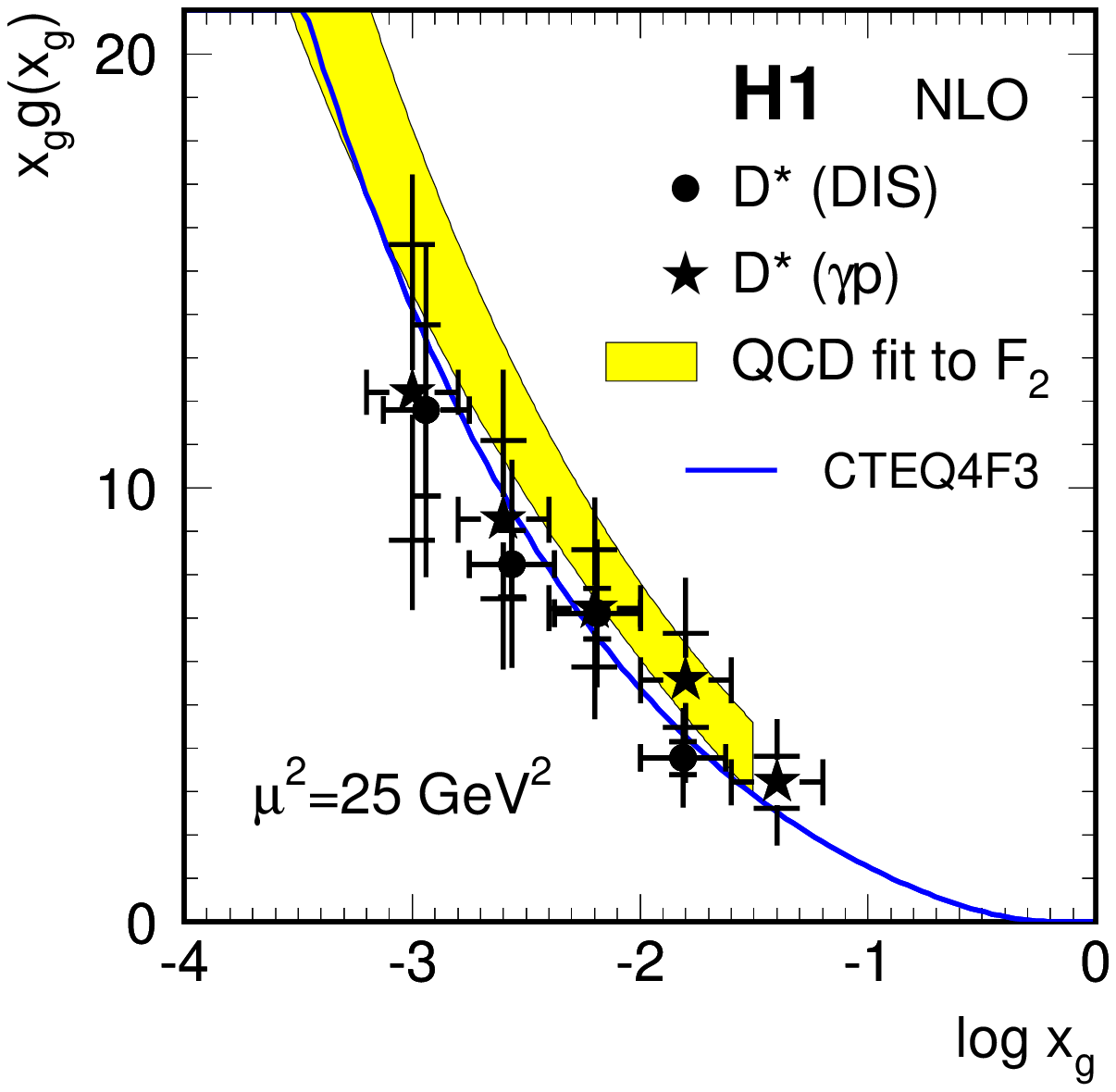,width=6.9cm}
\caption{\label{fig:glue} Gluon distribution.}
}
 
\subsection{Photo-production}

In the case of photo-production of open charm, additional, so-called
``resolved'' photon contributions have to be taken into account. 
The photon may fluctuate into a hadronic state, and a parton from 
this state interacts with a parton from the proton, e.g.\  
charm may be formed via 
gluon-gluon fusion. 
Since only a fraction of the photon's energy is available in the hard
sub-process, in comparison with ``direct'' photon interactions, 
the outgoing charm is found at lower transverse momenta and at 
more forward rapidities. 
Resolved photo-production is therefore suppressed by experimental cuts, 
and in fact extra care was taken in the above-mentioned gluon analysis to
minimize uncertainties related to such {\it a priori} unknown contributions.

There are two approaches in NLO QCD to describe charm photo-production. 
In the ``massive'' scheme~\cite{frixione}, 
charm is produced only dynamically in the 
final state (like in the DIS case above). 
In the ``massless'' scheme~\cite{kniehl}, charm also plays an active role 
in the initial state as a parton of the proton and of the photon. 
While the first approach should be adequate near threshold where 
the transverse momenta are comparable to the charm quark mass, the second 
has been developed for the region of higher $p_{\perp}$.
\FIGURE{
    \unitlength1cm
\begin{picture}(6.9,15)(0,0.3)
\put(0,9){\epsfig{file=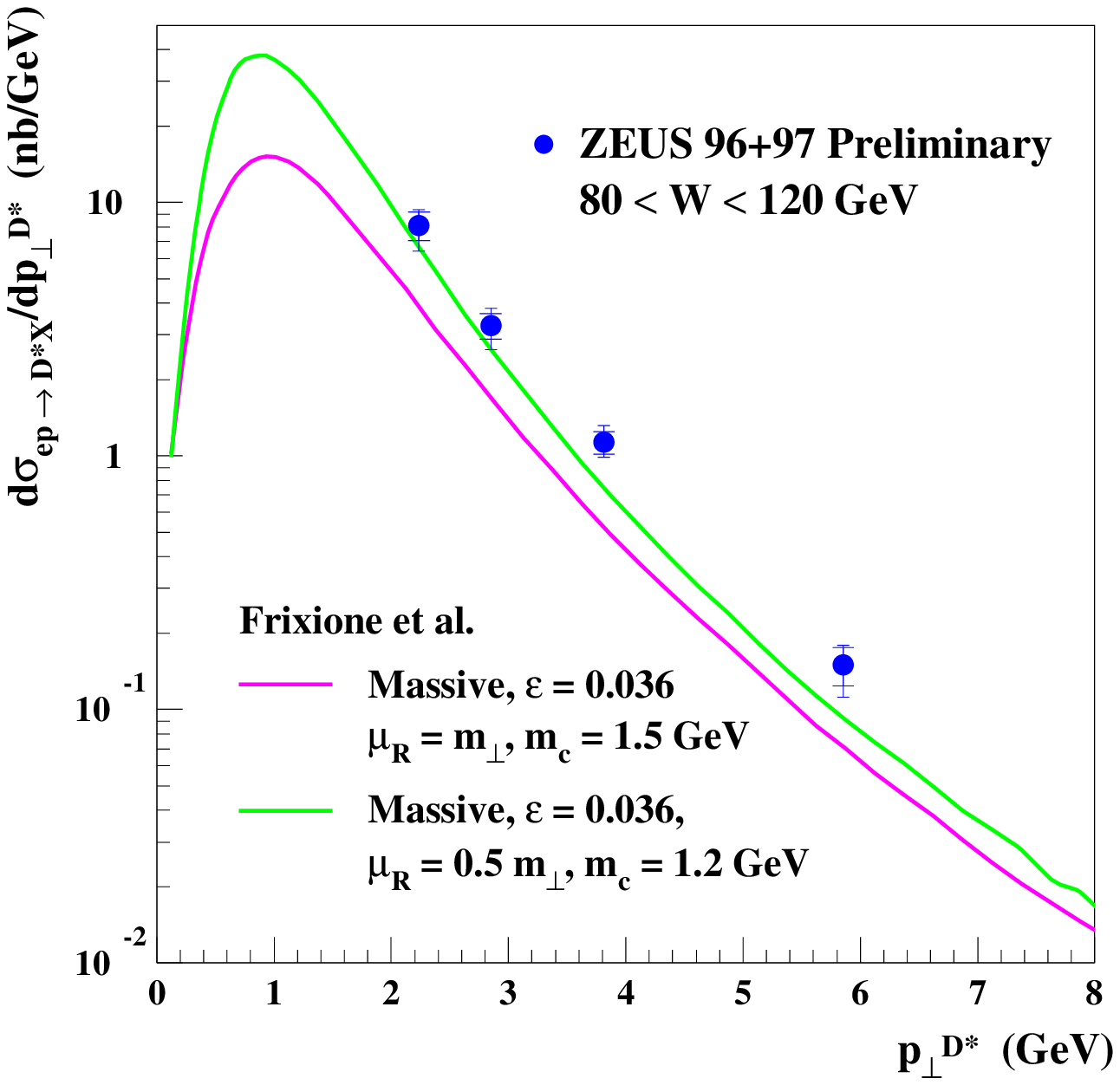,width=6.9cm}}
\put(0,4.5){\epsfig{file=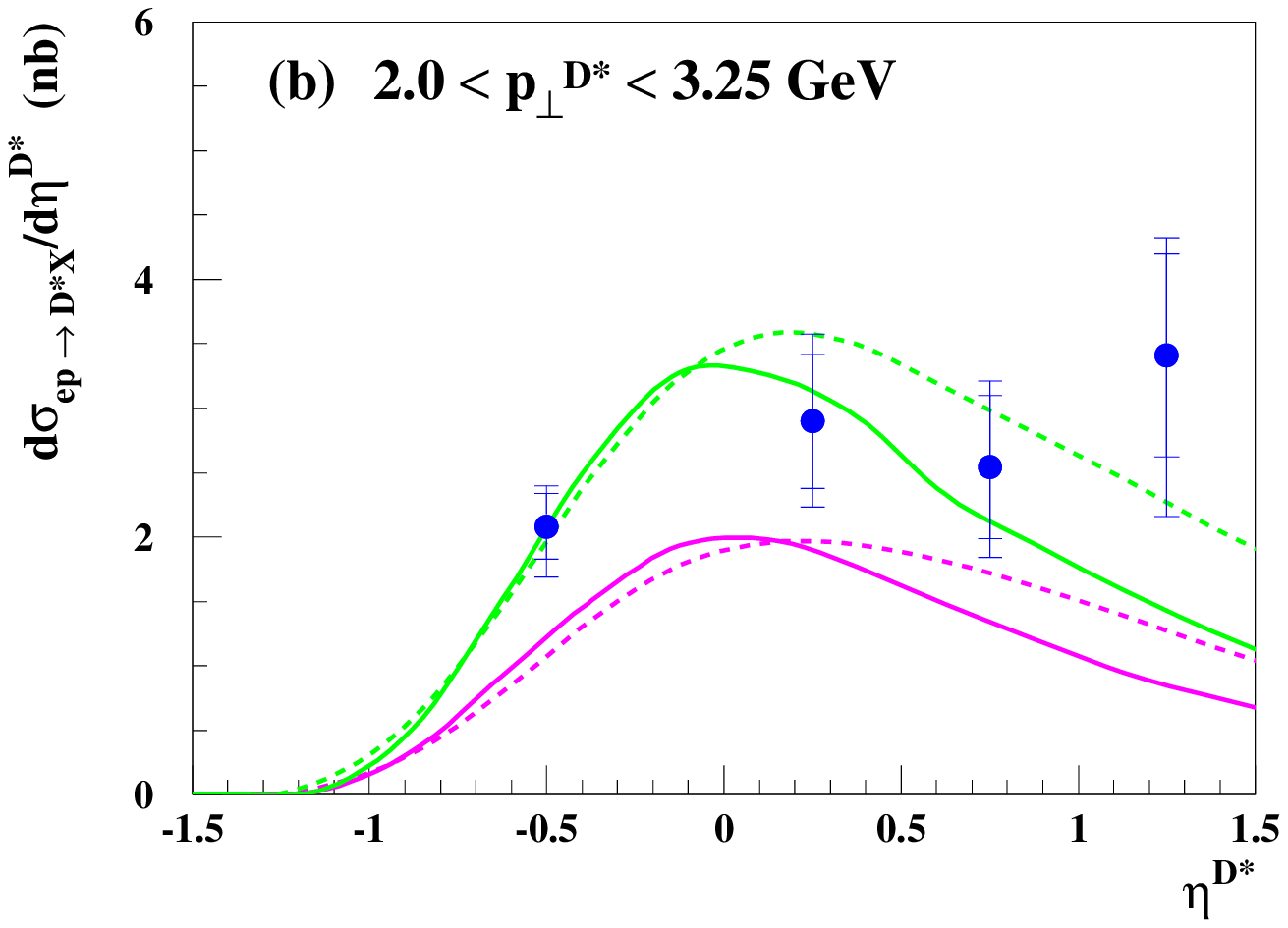,width=6.9cm}}
\put(0,0){\epsfig{file=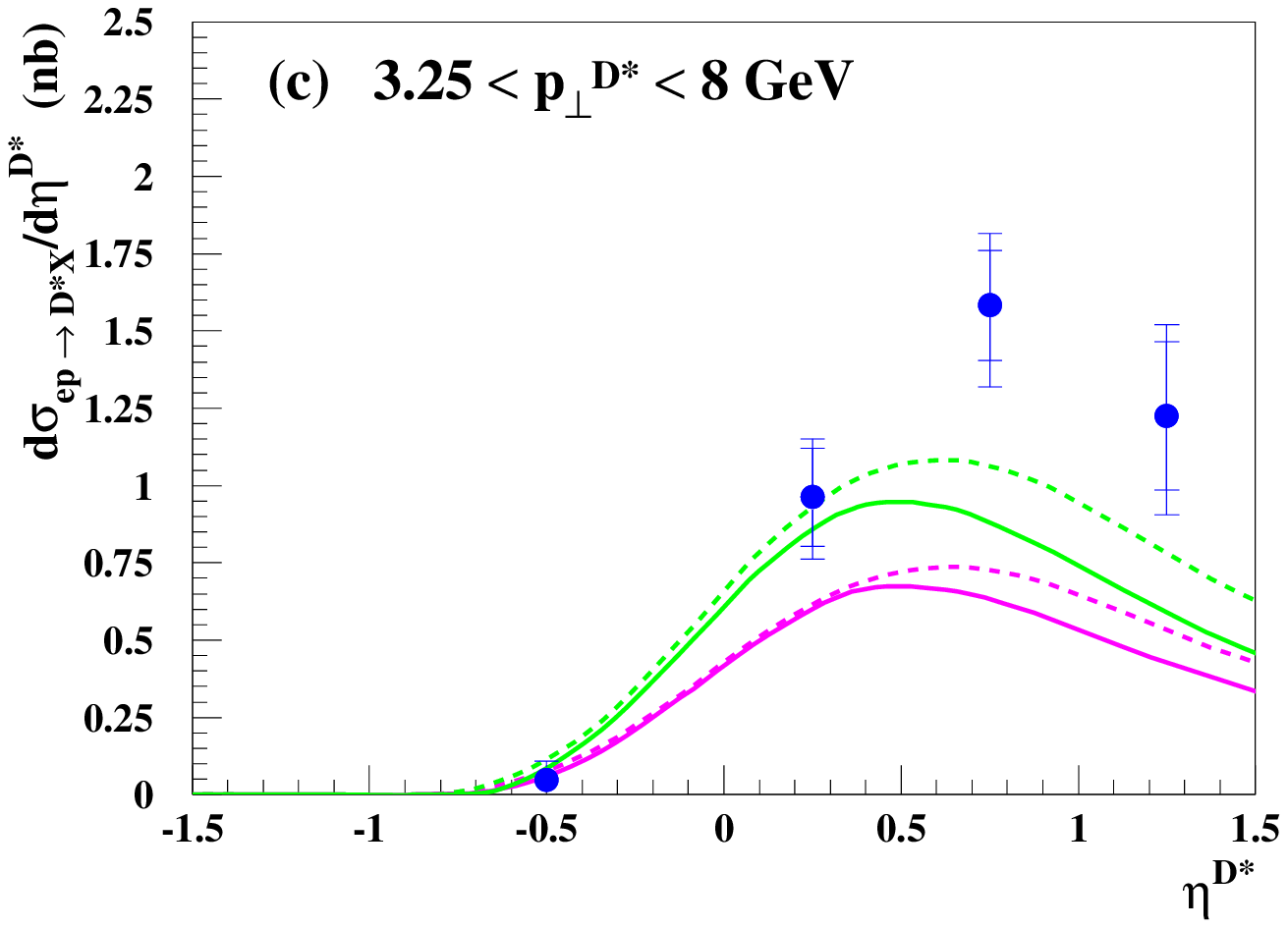,width=6.9cm}}
\end{picture}
\caption{\label{fig:dstgp} $D^{\ast}$ photoproduction cross section,
compared with NLO QCD in the ``massive'' scheme.
(Dashed: Peterson, dotted: PYTHIA fragmentation).}
}
 
$D^{\ast}$ photoproduction cross sections 
as recently measured by ZEUS~\cite{zeusdsubs}
in the region 80 $< W <$ 120 GeV  
are compared to NLO QCD calculations in the massive scheme in 
figure~\ref{fig:dstgp}. 
Fair agreement can be seen in the low $p_{\perp}$ range, with however
somewhat ``stretched'' parameter values. 
At higher $p_{\perp}$ the data tend to be above the prediction, 
in particular in the forward region.
As in the DIS case, predictions in this region are seen to be sensitive to 
details of how the fragmentation process is being modeled. 
The same conclusions can be drawn from the ZEUS data obtained with 
$D^{\ast}$ and $D_s$ mesons at higher $W$~\cite{dstgpzeushiw}. 
H1, albeit with somewhat larger measurement errors, do not observe 
a discrepancy with respect to the ``massive'' calculations~\cite{glue}. 

In the ``massless'' calculation,  
there is a large ``resolved'' 
contribution which is mostly due to diagrams where a charm quark 
from the photon interacts with a parton from the proton;
an example is shown in figure~\ref{fig:dstgpg}.
(Note that the assignment of ``direct'' and ``resolved''
is not unambiguous in NLO.) 
In this picture the cross sections are thus sensitive to the charm density
in the {\em photon},
as indicated in figure~\ref{fig:dstgpg},
where different photon PDFs have been used in the calculation.  
H1~\cite{glue} and 
ZEUS data~\cite{dstgpzeushiw} (shown here) 
have been compared with NLO QCD in the ``massless'' scheme. 
As in the ``massive'' case, 
the agreement is not good everywhere.
\FIGURE{
    \unitlength1cm
\epsfig{file=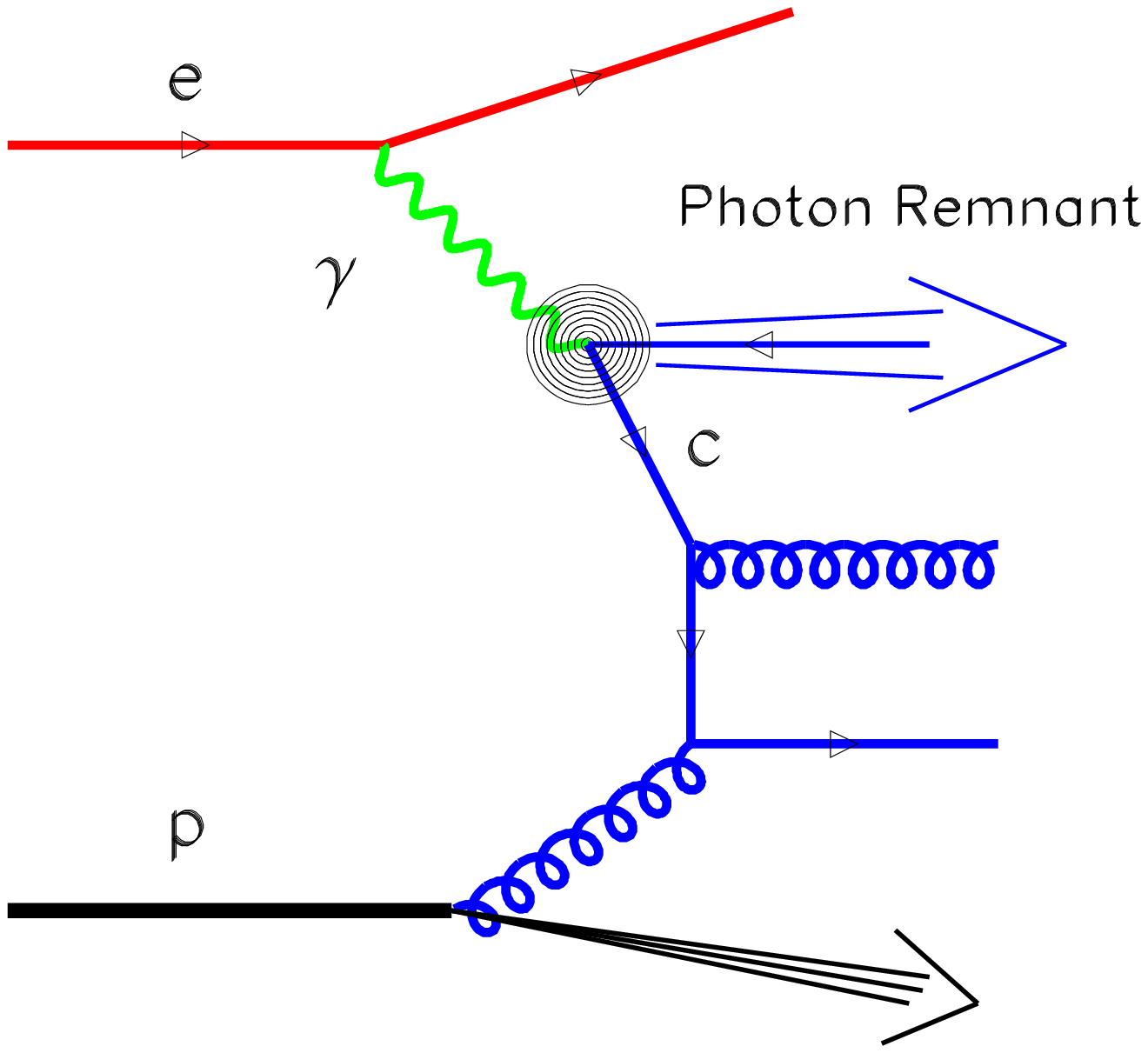,width=5cm}
\epsfig{file=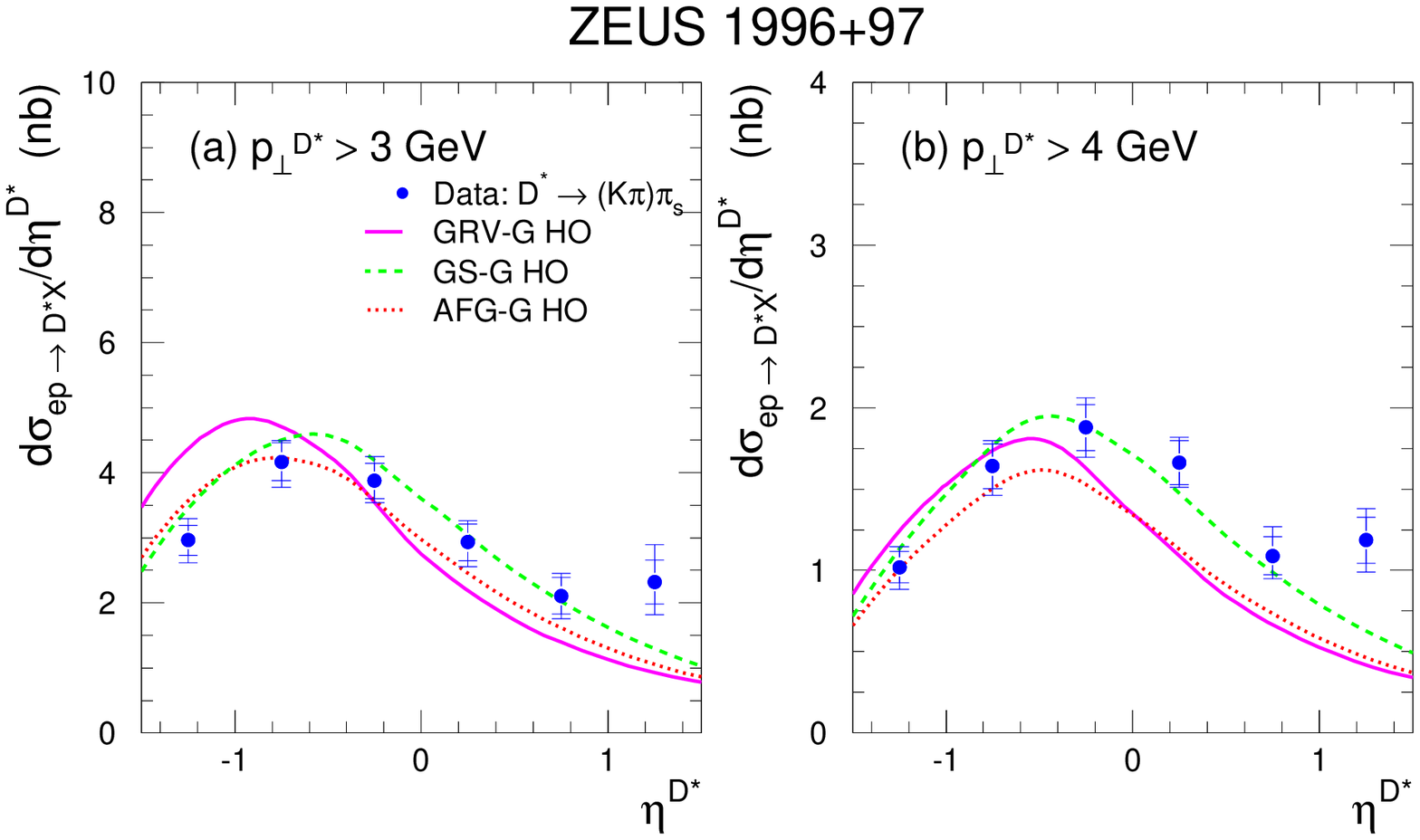,width=9cm}
\caption{\label{fig:dstgpg} $D^{\ast}$ photoproduction 
in the ``massless'' picture; $D^{\ast}$ cross sections 
compared with NLO QCD in the ``massless'' scheme, where the curves 
correspond to different photon PDFs.}
}

\subsection{Diffractive production}

Pursuing these concepts further, charm production may also shed
light on the dynamics of diffractive scattering.
A fraction of about 10 \% of the DIS events have a ``rapidity gap'': 
in contrast to the general case the region surrounding the outgoing 
proton beam is void of any particles. 
This distinctively ``diffractive'' topology leads to an interpretation of the 
events in terms of the exchange of a color-less object,
which may be identified with the Pomeron ($\pom$) 
in the framework of hadron-hadron interaction phenomenology. 
HERA offers the possibility to study these phenomena
in a perturbative regime and to investigate the partonic structure of the 
exchange, where
charm production again singles out the gluonic component.
\FIGURE{
   \unitlength0.72cm
   \begin{picture}(6.0,12)(1,4)
   \put(-2.0,8.1){\epsfig{width=8.4cm,file=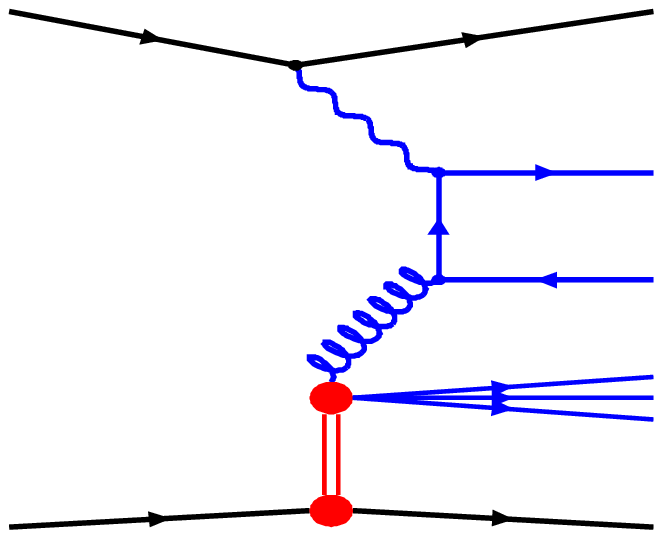}}
   \put(3.4,12.5){\small $g$}
   \put(0.5,9.9){\small $p$}
   \put(0.5,14.9){\small $e$}
   \put(7.2,9.9){\small $p'$}
   \put(7.0,14.9){\small $e'$}
   \put(6.8,14.0){\small $c$}
   \put(6.8,12.9){\small $\bar{c}$}
   \put(4.6,14.3){\small $\gamma$}
  \put(-.5,1.3){\epsfig{width=7.2cm,file=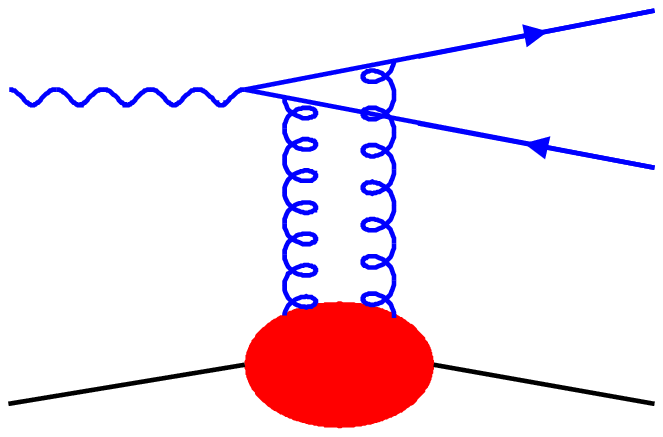}}
  \put(0.6,4.5){\small $p$}
  \put(7.8,4.5){\small $p'$}
  \put(7.9,8.3){\small $c$}
  \put(7.9,6.7){\small $\bar{c}$}
  \put(0.7,7.5){\small $\gamma$}
\end{picture}
\caption{\label{fig:ddmod} Diagrams of diffractive $D^{\ast}$ production:
``resolved Pomeron'' (upper) and ``2 gluon'' model.}
}

Two different model predictions are considered here
(figure~\ref{fig:ddmod}). 
In the ``resolved Pomeron'' model~\cite{respom}, 
the structure of the exchanged object is 
described in terms of quark and gluon densities which have been 
obtained from a fit to inclusive diffractive DIS data~\cite{f2d3}. 
In the ``two gluon'' model~\cite{two-g}, the color-less 
exchange is realized in terms of two hard gluons, 
and cross sections are sensitive to the gluon density in the proton.  
The two approaches lead to remarkably different kinematic distributions.
Both H1~\cite{shdis} and ZEUS~\cite{zeusddd} have presented first
measurements of differential cross sections this year.
\FIGURE{
    \unitlength1cm
\epsfig{width=6.5cm,file=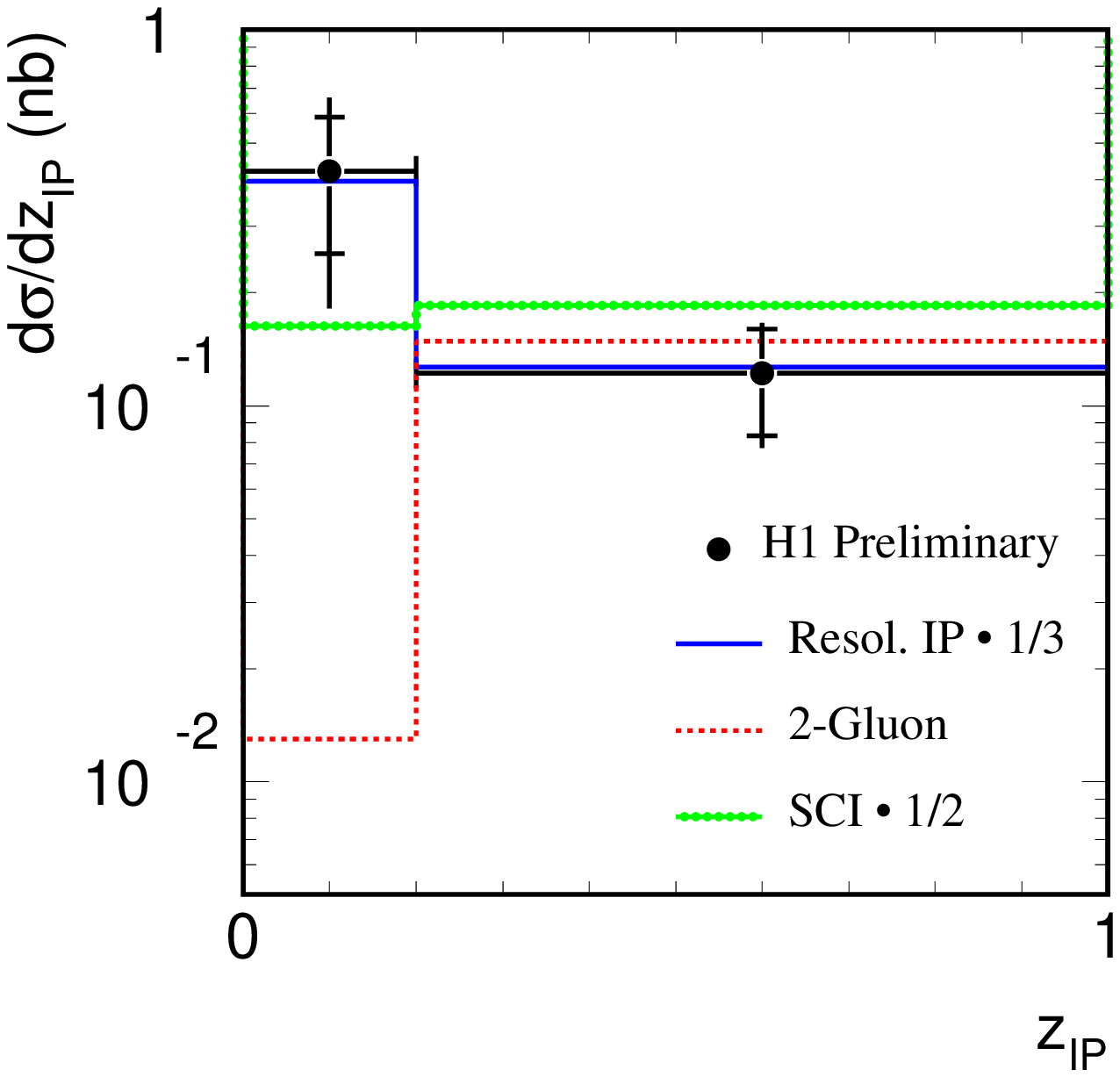}
\caption{\label{fig:difdif} $D^{\ast}$ cross section in diffractive DIS.}
}

$z_{\pom}$ is an observable correlated with 
the fraction of momentum of the exchanged object
which is carried by the parton interacting with 
the $\ccb$ pair.
In the ``$2 g$'' model, 
$z_{\pom}=1$ would hold if the partons could be directly observed.
The data~\cite{shdis} in figure~\ref{fig:difdif}
reveal that a sizeable fraction of charm is produced 
at low $\zpom^{obs}$ (``resolved Pomeron interaction''). 
The ``resolved $\pom$'' approach 
however  fails in reproducing the normalization of the H1 data. 
In the ``two gluon'' model, on the other hand, 
a lack of higher order, low $z_{\pom}$ 
contributions is apparent. 
More general, models, where the hadronic system $X$ 
predominantly consists of the $c\bar{c}$ system alone, are disfavored.
The data still have large, mostly statistical, uncertainties, 
but they already provide valuable and rather distinctive 
indications for further refinement of the theoretical description.

\section{Beauty Production}
Finding beauty at HERA is not straightforward. 
Theory predicts total cross section ratios of about 
$\sigma_{\rm uds}:\sigma_{\rm charm}:\sigma_{\rm beauty}\sim2000:200:1$~\cite{frixione}.
The measurements to date rely on the well established 
signature of semileptonic decays of $b$ hadrons in jets. 
Due to the higher $b$ mass, the leptons in $b$ events have 
a higher transverse momentum $p_T^{rel}$ with respect to the jet direction 
which approximates the flight direction of the decaying hadron
(figure~\ref{fig:bptrel}).
The $b$ cross section can thus be extracted by means of a fit to the 
$p_T^{rel}$ distribution. 
\FIGURE{
    \unitlength1cm
\begin{picture}(15,8)
\put(0.5,1.5){\epsfig{file=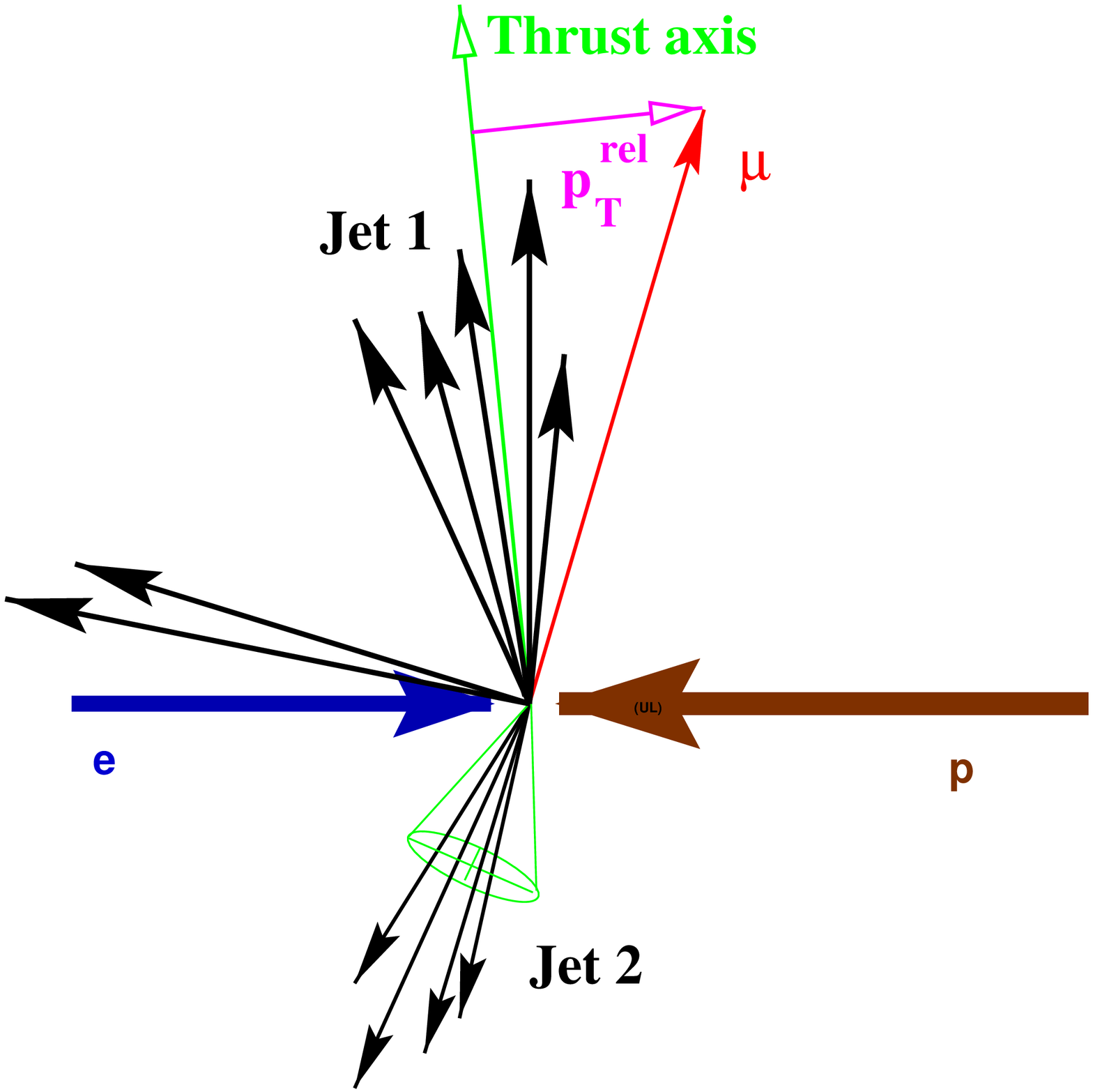,width=7cm}}
\put(7.5,0){\epsfig{file=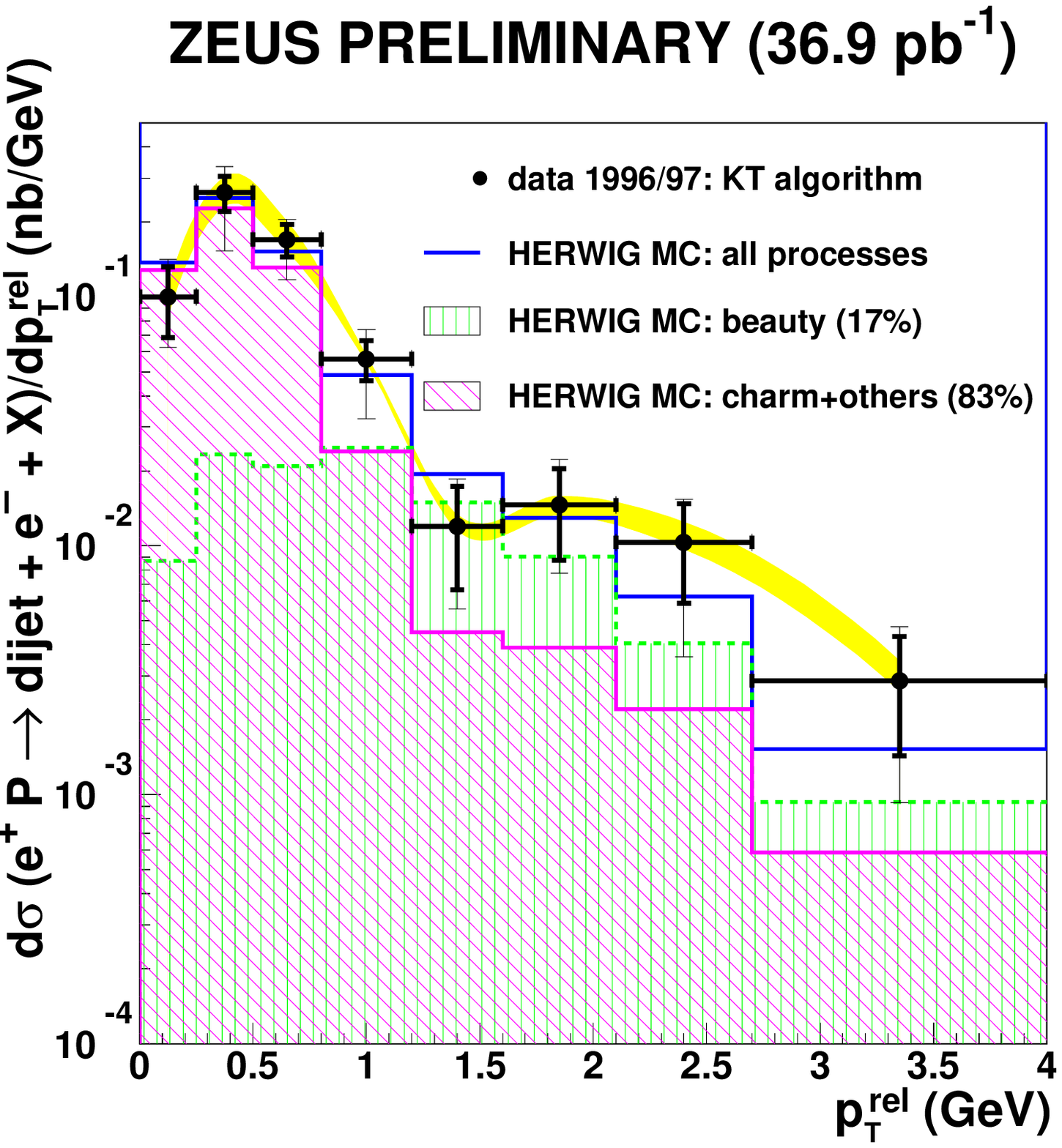,width=8cm}}
\end{picture}
\caption{\label{fig:bptrel} $b$ production signature and 
$p_T^{rel}$ spectrum.}
}

The preliminary H1 result released in 1998 has meanwhile been 
published~\cite{openb}.
The visible cross section, quoted for the range 
$Q^2<1\,{\rm GeV^2}$,  $0.1<y<0.8$, $p_t^\mu>2\,{\rm GeV}$,
$35^\circ<\theta^\mu<130^\circ$ 
of 
$\sigma^{vis}_{b\bar{b}} = 0.93\pm0.08^{+0.17}_{-0.07}\,{\rm nb}$
is found to be surprisingly high:
the Monte Carlo prediction based on Leading Order QCD calculation
is 4 to 5 times lower ($\sigma^{vis}_{b\bar{b}} = 0.191\,{\rm nb}$).
NLO corrections are however non-negligible, as in the charm photoproduction 
case.

\FIGURE{
    \unitlength1cm
\epsfig{file=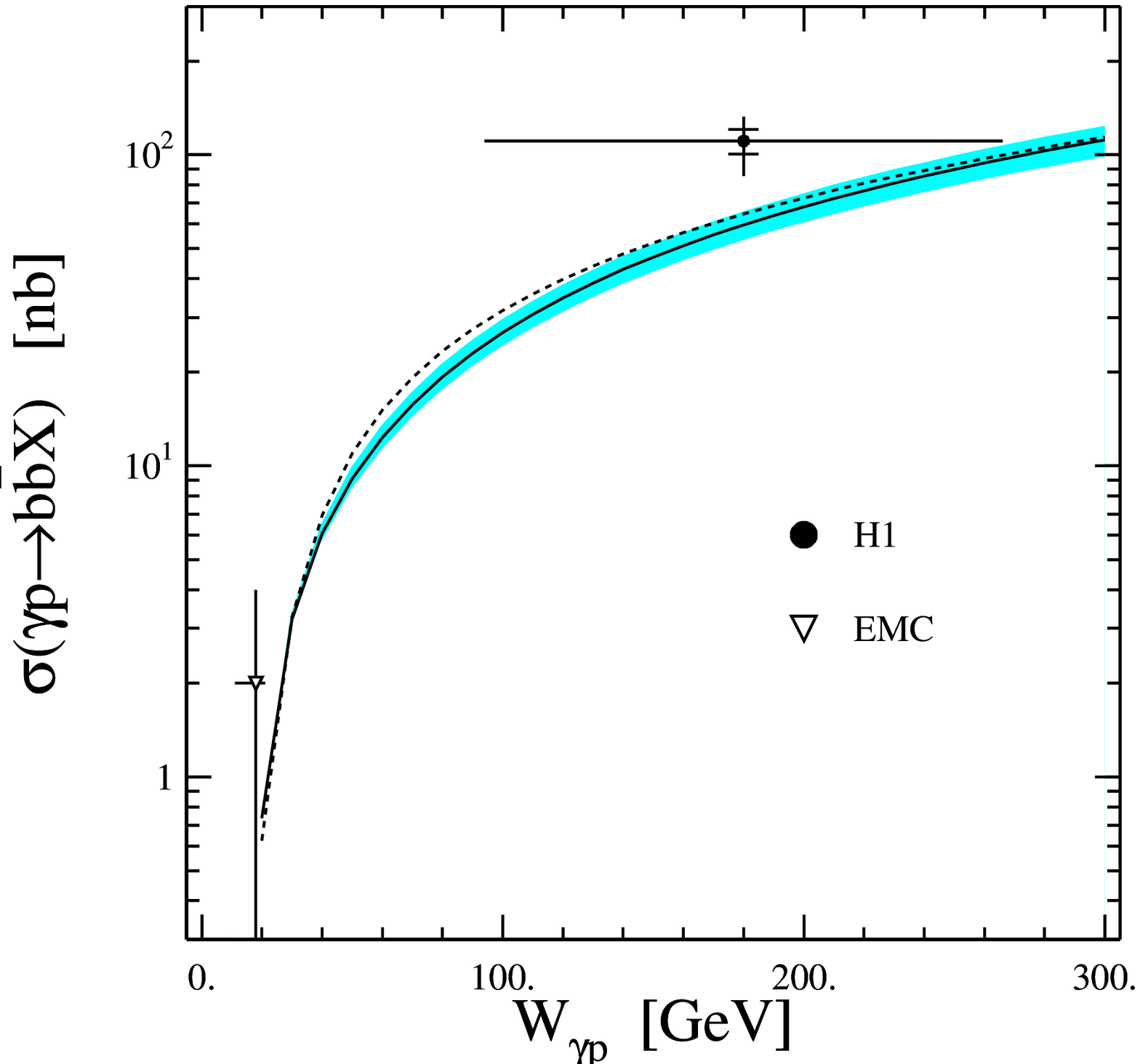,width=7cm}
\epsfig{file=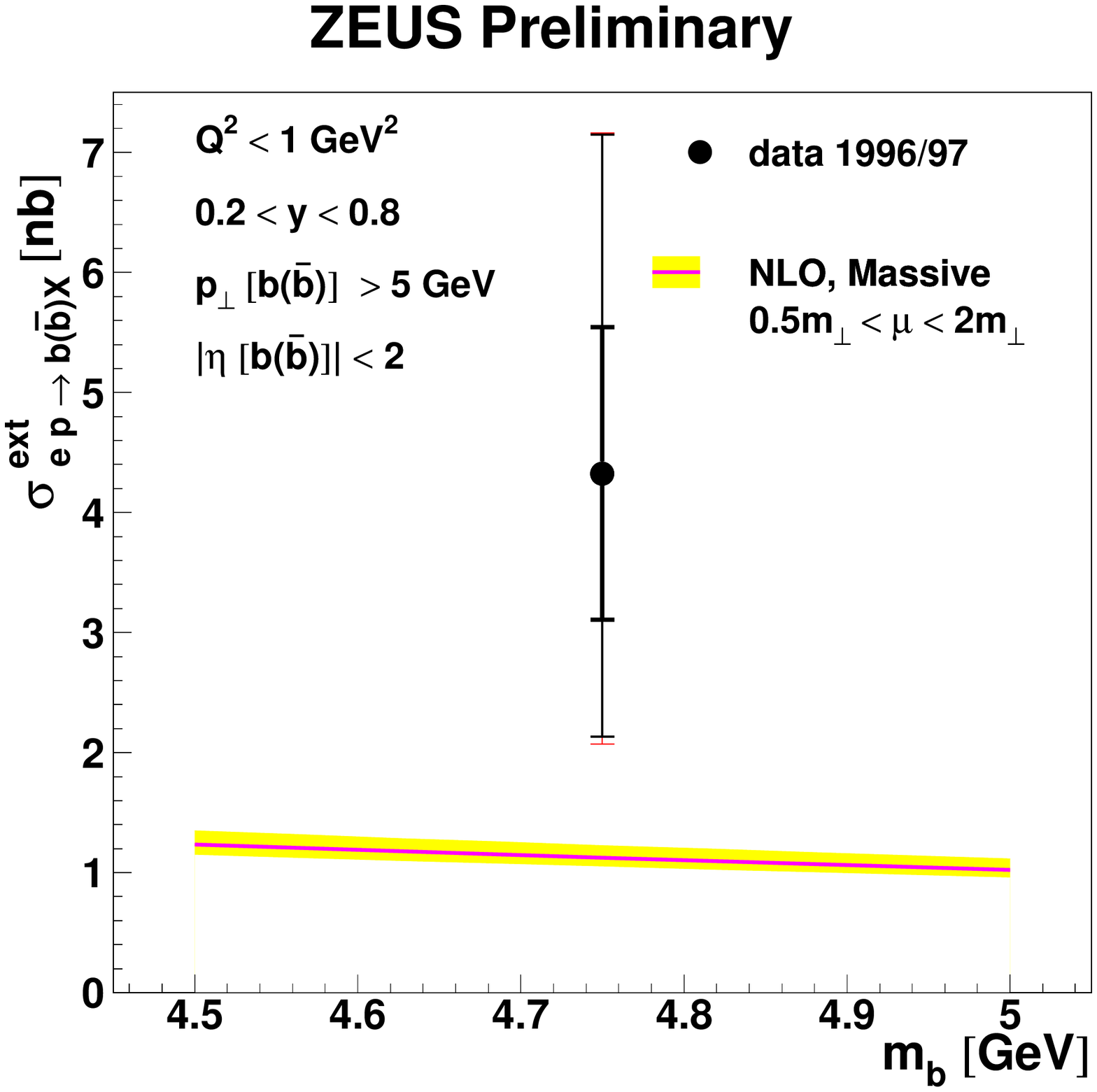,width=6.9cm}
\caption{\label{fig:bnlo} $b$ production cross section, compared with NLO QCD.}
}
In figure~\ref{fig:bnlo} the result, extrapolated to the full phase space by 
means of a NLO program incorporating a Peterson type model of fragmentation, 
is confronted with the NLO prediction. 
Theory at NLO still undershoots the experimental result
by about a factor of 2.
This is in marked contrast to the expectation that predictions for beauty 
should be more reliable than for charm,  
due to the higher scale set by the $b$ mass. 
On the other hand, the NLO approximation
also has difficulties reproducing the normalization
of open $b$ cross section measurements at the Tevatron~\cite{openbattev}.
  
ZEUS has determined the beauty cross section in three independent analyses, 
using electrons, muons in the central region 
and muons in the forward detector region~\cite{zeusb}.
The $p_T^{rel}$ distribution, obtained in the $e$ channel, is displayed in 
figure~\ref{fig:bptrel}.
Overlayed is a Monte Carlo prediction obtained with the HERWIG generator, 
which already describes the shape of the data reasonably well.
It should be noted that in HERWIG about half of the $c$ and $b$ 
photo-production is due to resolved processes with a heavy quark 
in the initial state, mostly as an active parton in the photon. 
A fit of the different components yields a $b$ fraction of 
20\%, which translates into a visible cross section of  
$\sigma_{vis}(e^+\ra\mbox{\rm 2 jets} + e^- + X)
              = 39 \pm 11 ^{+23}_{-16}$ pb
for events with 
2 jets ($E_T^{1,2} > 7,6$ GeV, $|\eta | < 2.4$)
in the kinematic range  
$Q^2<1$GeV$^2$, 0.2$<y<$0.8, $e$: $P_T>1.6$GeV, 
              $|\eta|<1.6$.

Consistent results are obtained in the muon channels, 
with the same jet requirements and for the same kinematic range. 
For central muons ($-1.75<\eta<$1.3, $p>$3 GeV)
$\sigma_{vis}(e^+\ra\mbox{\rm 2 jets} + \mu + X)
              = 36.4 \pm 5.2 ^{+10.4}_{-9.1}$ pb
is quoted, and for forward muons (1.4$<\eta<$2.4, $p_{\perp}>$1.5 GeV)
$\sigma_{vis}(e^+\ra\mbox{\rm 2 jets} + \mu + X)
              = 20.5 \pm 6.5 ^{+6.3}_{-5.7}$ pb.

Using the Monte Carlo to convert the measured hadronic cross section 
to a partonic cross section 
allows a comparison with NLO QCD calculations at parton level. 
This is done for the ZEUS result in the $e$ channel in figure~\ref{fig:bnlo}. 
Here, the result is found to be even a factor of 4 above theory, 
with somewhat larger errors.   

\FIGURE{
\unitlength1cm
  \begin{picture}(14,8)(0,0)
\put(-1,0){\epsfig{file=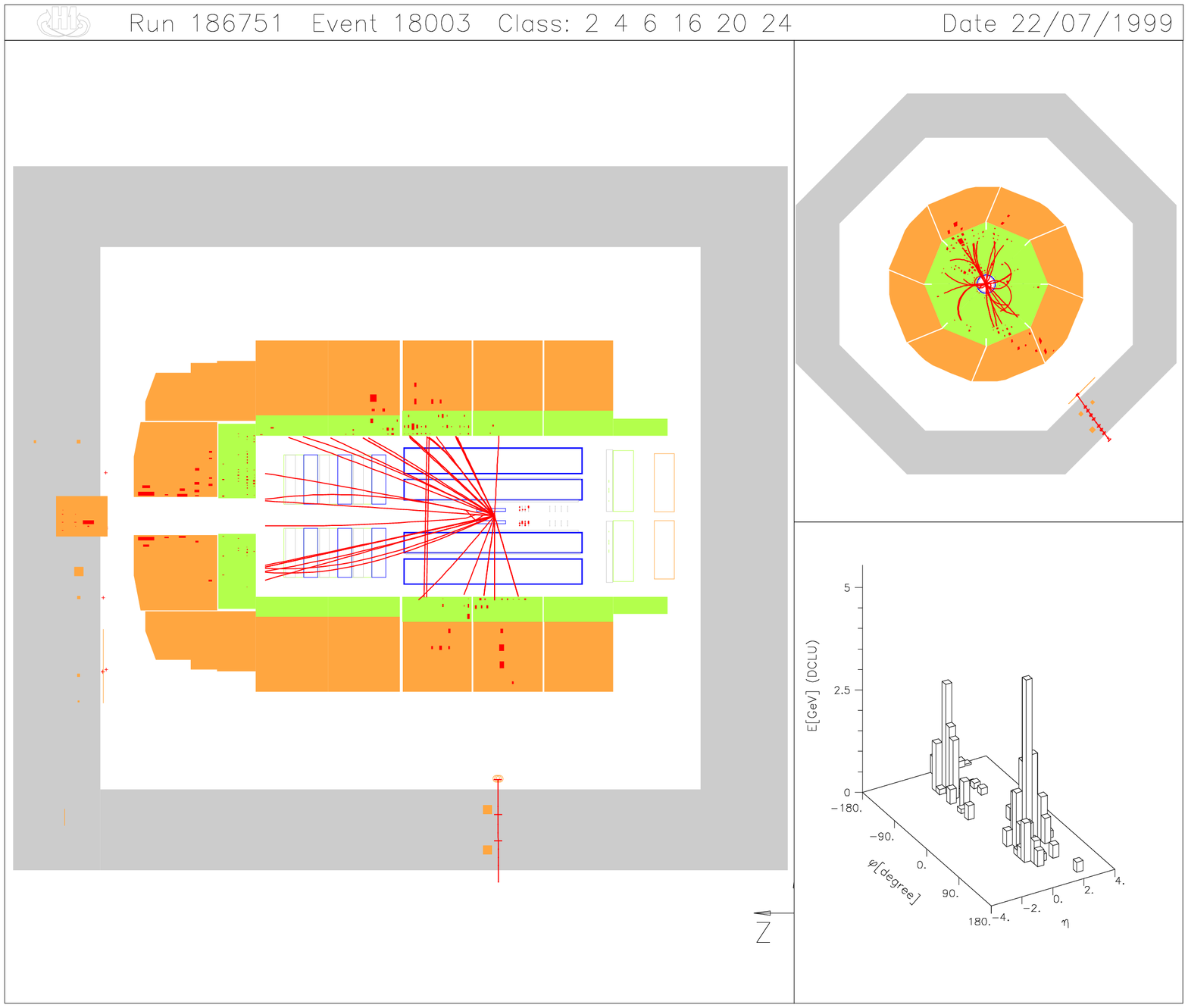,
width=7.5cm,angle=90}}
\put(8,0.1){\epsfig{file=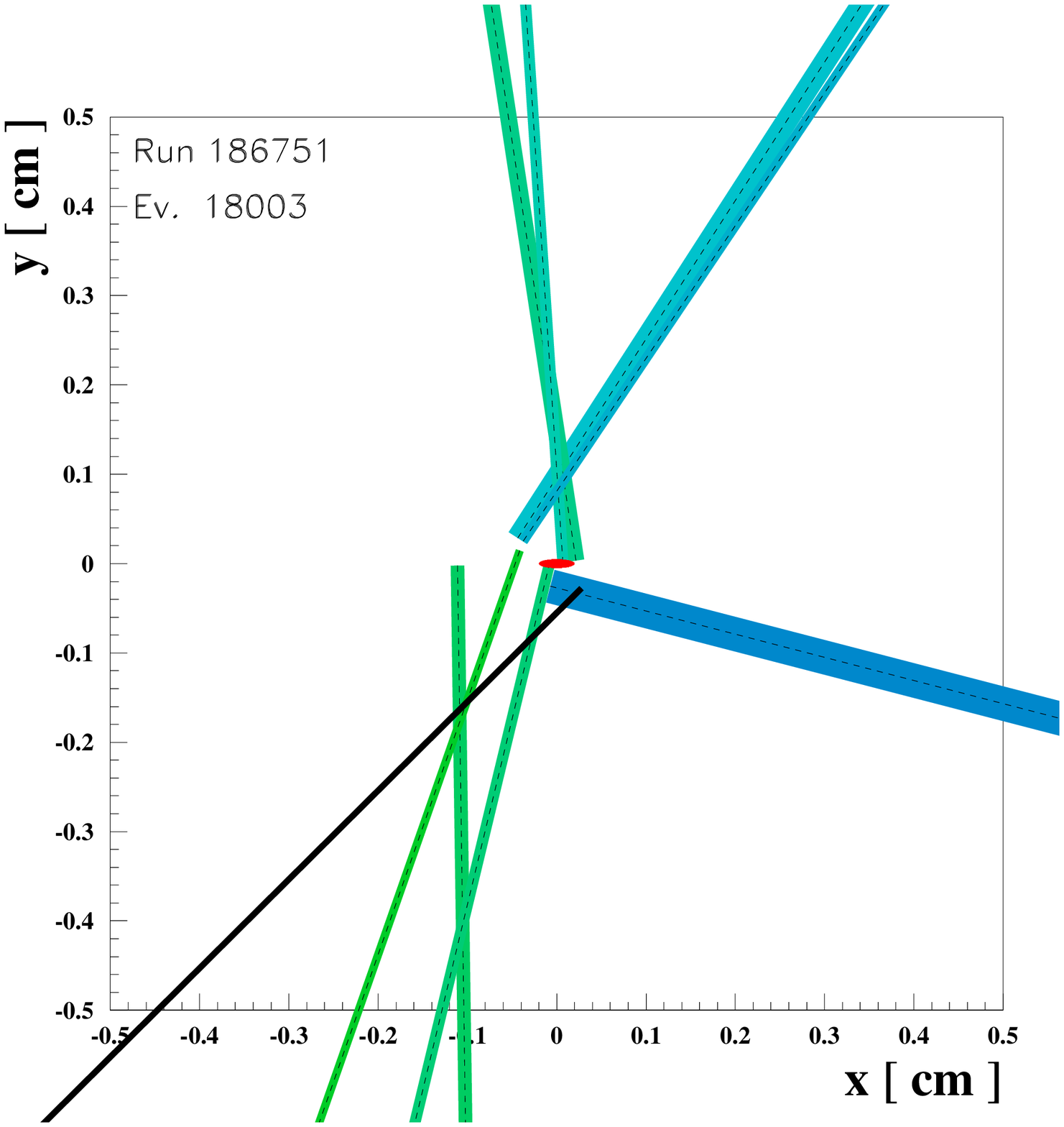,
width=6.5cm}}
\put(9.5,2){\Large\red $\mu$}
\put(12,4){\normalsize\sf beam spot} 
\end{picture}
\caption{\label{fig:bcand} A $b$ candidate in the H1 detector;
right: vertex region.}
}
The long lifetimes of $c$ and $b$ hadrons which can be measured 
with microvertex detectors provide an independent signature.
With its Central Silicon Tracker (CST), H1 has for the first time  
at HERA attempted to exploit these features. 
An event with two jets and a penetrating track identified as a muon 
is shown on the left-hand side of figure~\ref{fig:bcand}.
The muon has a relatively high
transverse momentum of 3 GeV$/c$ with respect to the nearest jet, 
which is unlikely for muons from charm decays or mis-identified hadrons   
but is expected for decays of $b$ flavoured hadrons.
In the magnified view perpendicular to the beam (on the right-hand side) 
the tracks measured in the CST are represented as bands with widths 
corresponding to their $\pm 1\sigma$ measurement precision.   
The resolution provided by the CST reveals that 
the muon track originates from a well separated secondary vertex.   
This is now being integrated into ongoing 
physics analyses. 

\section{Summary and Outlook}

Heavy quarks have become an increasingly interesting part
of the physics field opened up by HERA. 
With the statistical power now reached in the data, charm quarks 
are a direct probe of the gluonic component of hadronic structure. 
In the case of the proton, the fact that about one quarter of the $ep$ 
interactions in the HERA regime result in final states with charm 
makes it evident that understanding the charm contribution
to the structure function $F_2$ is a {\it conditio sine qua non}.
The agreement of the gluon density directly determined from charm   
with the result from the scaling violation analysis  
lends strong support to the underlying QCD picture. 

New light has also been shed on the structure of the photon, 
where a charm content in the ``resolved'' photon 
may provide a viable route to understanding the HERA data.  
Finally, probing the structure of diffractive exchange 
with charm, differential cross section data have started to discriminate 
between different theoretical concepts.

Beauty production at HERA has now been measured by both H1 and ZEUS, and 
found to occur at a rate that is presenting NLO QCD with a challenge.   
Although experimental errors are still considerable, a re-evaluation
of theoretical uncertainties might be appropriate.

Top pair production is kinematically ex\-clu\-ded at HERA energies. 
It has however been speculated~\cite{fritzsch} recently 
that the events with isolated leptons and large missing $p_{\perp}$,
of which the H1 collaboration~\cite{hiptlep} observes more than expected
(and some with atypical kinematics), 
might be due to single top production mediated by a new, 
flavour-changing effective interaction. 
The top quark would decay via $t\ra Wb$, $W\ra \mu\nu$,
giving rise to the observed signature. 
In fact, some of the observed outstanding events are 
kinematically not inconsistent with such a hypothesis. 
Lifetime-based tagging techniques will in the future be used to 
investigate such exciting possibilities further. 
One more reason why H1 and ZEUS physicists 
look forward to the high luminosity running 
at the upgraded HERA machine in 2001. 

\acknowledgments
I like to thank my colleagues in H1 and ZEUS for providing me with 
the results of their work for this presentation. 
And it is a pleasure to thank Paul Dauncey, Jonathan Flynn, and their crew
for organizing a top meeting at a beautiful place in such a
charming way.


\begin{thebibliography}{99}
\bibitem{naroska}
P. Merkel, talk at HEP 99, Tampere, 1999; \\
A. Bertolin, {\it ibid.}
\bibitem{lq}
H1 Coll.,
{\it A Search for Leptoquark Bosons and Lepton Flavor Violation in Positron-Proton Collisions at HERA},
DESY 99-081. 
\bibitem{zeusdsubs}
ZEUS Coll., contrib.\ paper No.\ 525 for HEP 99, Tampere, 1999. 
\bibitem{oldcharmglue}
NMC Coll., D.Allasia {\it et al.}, \plb{258}{1991}{493}.
\bibitem{frixione}
S. Frixione, M.L. Mangano, P. Nason, and G. Ridolfi, 
\plb{308}{1993}{137}, \plb{348}{1995}{633}; \\
S. Frixione, P. Nason, and G. Ridolfi, \npb{454}{1995}{3}.
\bibitem{disnlo}
B.W. Harris and J. Smith, \prd{57}{1998}{2806}.
\bibitem{peterson}
C. Peterson {\it et al.}, \prd{27}{1983}{105}.
\bibitem{opal0.035}
OPAL Coll., R.~Akers {\it et al.}, \zpc{67}{1995}{27}.
\bibitem{glue}
H1 Coll., C. Adloff {\it et al.}, \npb{545}{1999}{21}.
\bibitem{zeusf2c}
ZEUS Coll., 
{\it Measurement of $D^{\ast\pm}$ production and the charm contribution 
to $F_2$ in deep inelastic scattering at HERA},
DESY 99-101.
\bibitem{norrbin}
E. Norrbin, T. Sj\"ostrand, \hepph{9905493}.
\bibitem{kniehl}
J. Binnewies, B.A. Kniehl, G. Kramer, 
\zpc{76}{1997}{677}, \prd{58}{1998}{014014}
\bibitem{dstgpzeushiw}
ZEUS Coll., J. Breitweg {\it et al.}, 
{\em Eur.\, Phys.\, J.} {\bf C6} (1999) 67.
\bibitem{respom}
G. Ingelmann, P. Schlein, \plb{152}{1995}{256}.
\bibitem{f2d3}
H1 Coll., C. Adloff {\it et al.}, \zpc{76}{1997}{613}
\bibitem{two-g}
J. Bartels, H. Lotter, M. W\"usthoff,\plb{379}{1996}{239}.
\bibitem{shdis}
S. Hengstmann, \npps{79}{1999}{296}.
\bibitem{zeusddd}
ZEUS Coll., contrib.\  paper No.\ 527 for HEP 99, Tampere, 1999. 
\bibitem{openb}
H1 Coll., 
{\it Measurement of Open Beauty Production at HERA},
DESY 99-126.
\bibitem{openbattev}
CDF Coll., F. Abe {\it et al.}, \prl{71}{1993}{2396},\prd{53}{1996}{1051}; \\
D0 Coll., S. Abachi {\it et al.}, \prl{74}{1995}{3548}, \plb{370}{1996}{239}.
\bibitem{zeusb}
ZEUS Coll., contrib.\  paper No.\ 498 for HEP 99, Tampere, 1999. 
\bibitem{fritzsch}
H. Fritzsch, D. Holtmannsp\"otter, \plb{457}{1999}{186}.
\bibitem{hiptlep}
H1 Coll., C. Adloff {\it et al.}, {\em Eur.\, Phys.\, J.} {\bf C5} (1998) 575.
\end{thebibliography}
\end{document}